\documentstyle[colordvi,psfig,12pt]{article}

\begin{document}

\title{The inner structure of Zeldovich' pancakes}

\author{E. Aurell $^{1,2,3}$\thanks{e-mail: eaurell@sics.se}, D. Fanelli $^{3,4}$\thanks{e-mail: fanelli@et3.cmb.ki.se}, S.N. Gurbatov $^{5,6}$
\thanks{e-mail: gurb@rf.unn.ru}\,\, \& A.Yu. Moshkov $^5$\thanks{e-mail: moshkov@rf.unn.ru}}
\maketitle

\begin{center}
\begin{tabular}{ll}
$^1$ & SICS, Box 1263, SE-164 29 Kista, Sweden\\
$^2$ &  NORDITA, Blegdamsvej 17, DK-2100 Copenhagen, Denmark\\
$^3$ & Department of Numerical Analysis and Computer Science, KTH, \\
& S-100 44 Stockholm, Sweden\\
$^4$ & Department of Cell and Molecular Biology, Karolinska Institute, \\
& SE-17177 Stockholm, Sweden \\
$^5$ & Radiophysics Department, University of Nizhny Novgorod,\\
& 603600, Gagarina Av. 23, Russia, (permanent address) \\
$^6$ & Observatoire de la C\^ote d'Azur, Lab. G.D.~Cassini,\\
& B.P.~4229, F-06304 Nice Cedex 4, France.\\
\end{tabular}
\end{center}

\begin{abstract}
The evolution of a planar perturbation in a Ein\-stein-de~Sitter
Universe is studied using a previously introduced Lagrangian scheme.
An approximate discrete dynamical system is derived,
which describes the mass agglomeration process qualitatively.
Quantitative predictions
for the late density profile are obtained 
therefrom, and validated by numerical
simulations. 
The main result is a scaling regime for the density 
profile of a collapsing object of mass $M$
around cosmological coordinate
$r^*$, $\rho(r)\sim \frac{M}{d}\left(\frac{|r-r^*|}{d}\right)^{-\frac{1}{4}}$.
The characteristic scale of
the agglomeration, $d\sim \left(t/t_0\right)^{\frac{4}{9}}$,
is an increasing function of cosmological time $t$. 
The major part of the mass hence always 
lies in a region with decreasing mass density. This shows that
one-dimensional self-gravitating motion is not sufficient to 
effectively drive structure formation in an Ein\-stein-de~Sitter Universe.
These results are compared with analogous investigations
for the adhesion model (Burgers equation with positive viscosity),
where the agglomeration is faster, and one-dimensional dynamics
is effective.
We further study the mutual motion of two mass agglomerations, and
show that they oscillate around each other for long times, 
like two ``heavy particles''.
Individual particles in the two agglomerations do not mix
effectively on the time scale 
of the interagglomeration motion.
\end{abstract}

\begin{flushleft}
PACS numbers\,: 02.50.Ey, 
                       05.60.-k, 

{\it Keywords:} Self-gravitating dynamics; Mass density; Adhesion model; Burgers equation

\end{flushleft}



\section{Introduction}
\label{s:intro}
Structure formation in the Universe is a rich and challenging
problem, which touches many sides of Physics.
It is generally believed that the presently observed large scale
structures have been generated by
the process of the gravitational instability, acting on initially
small density perturbations.
The temperature fluctuations on the uniform Cosmic Microwave Background
(CMB) radiation provide an image of these fluctuations,
which were the seeds of the present large scale distribution of matter
\cite{COBE-web-site,DMR-web-site,BOOMERANG}.
The observed CMB angular power spectrum is dominated by a peak at $1$ degree
of arc,
and also shows structures at smaller scales.
\\\\
According to present consensus,
the constituents of our Universe are ordinary
matter (5\%), dark matter (25\%) and dark energy (70\%).
The three forms can be summarily classified by how their energy
densities change with a
cosmic scale factor $a$: ordinary and dark matter behave as $a^{-3}$,
radiation, which interacts with ordinary matter, behaves as $a^{-4}$,
while dark energy, at least in the simplest models thereof,
is independent of $a$.
Hence, even if dark energy dominates today, dark matter and ordinary matter
dominated earlier.
Further, the Universe is flat, 
in agreement with an early epoch of inflation.
Within that scenario, the primordial density perturbations were
random Gaussian with power spectrum $P(k) \sim A k~$, which is also
consistent with the data.
The ordinary matter content
can be deduced from observations at e.g. visible or radio frequencies.
The presence of dark
matter can be inferred from the observed dynamics of cosmic objects,
particularly from fast rotation of hydrogen
clouds far outside the luminous disc of spiral galaxies, as well as
high-velocity dispersion of galaxies in clusters~\cite{Weinberg1972}.
Dark energy, or quintessence~\cite{cal}, is equivalent
to a non-zero cosmological constant, $\Lambda$, in Einstein's
equations~\cite{Weinberg1972}, and there is recent support for a non-zero
$\Lambda$ also from redshift observations.
All of these direct measurements can be 
compared with theoretical cosmology
and the observed angular structure of the CMB.
A sequence of peaks
should indeed arise
from coherent acoustic oscillations in the baryon-photon 
fluid during an early epoch.
Their amplitudes and relative 
positions provide another series of tests of  
cosmological models, and put a different
series of constraints on the parameters
of such models, c.f.~\cite{Tegmark-science}
for a recent review.
\\\\
The growth of small initial perturbations to large-scale structures
is essentially a problem in classical gravitational
physics, where all the above
mainly enters in the initial conditions, and the how the Universe
expands as a whole. The problem is nevertheless
far from trivial. 
A principal
difficulty is that a self-gravitational
medium has no ground-state, and the dynamics is not ergodic.
Contrary to e.g. molecular dynamics, one cannot appeal to
a shadowing lemma, and argue that even if a simulation only
solves the equations of motion in a rough approximation,
the simulated system nevertheless stays close to the real system,
with different initial conditions. The single most influential
result in structure formation, Zeldovich' pancake theory, was indeed
developed without any recourse to simulation at all, but purely
by extending linear analysis into the non-linear domain. The fact
that this model and the adhesion model are still the benchmarks
today shows the difficulty of firmly establishing the detailed
characteristics of the mass agglomeration process.
\\\\
The goal of this paper is to investigate the non-linear regime of the
gravitational instability in the special case that the initial
perturbations are planar. We also assume an Einstein-de Sitter Universe,
and hence disregard the dark energy component.
With these limitations, one can introduce a Lagrangian integration scheme,
called the Quintic (Q) model~\cite{Qmodel,fanelli-aurell}, which is
both fast, and exact from collision to collision. 
It is worth stressing that the Q model is just one of the possible
representation of the dynamics \cite{fanelli-aurell}. A conceptually
equivalent formulation was previously proposed by \cite{YANO-GOUDA}. These two
approaches share the characteristic that they allow to avoid sources
of additional errors, like truncation, involved in any approximate
numerical scheme (see \cite{DOROSHKEVICH}). 

The focus is here on the inner structure of a pancake. The novelty with
respect previous investigations \cite{DOROSHKEVICH,YANO-GOUDA} is that,
by making use of a discrete map to approximate the exact Q dynamics, we
derive a close analytical expression for the density profile. These results 
are validated by direct
numerical simulation, and compared with analogous results for the
adhesion model.
We further investigate the pair-wise motion of two pancakes under
their mutual attraction.
\\\\
The paper is organized as follows: in Section \ref{s:VP} we present the
general background, and in Section \ref{Q} we re-derive, for
completeness, the Q model.
In Section \ref{s:2party} the evolution of a two particle system is
considered: an approximate discrete system is derived and validated
by numerical investigations. The analysis is extended in
Section \ref{s:spatialdens} where the inner structure of a single cluster
is investigated. Improving previous results in~\cite{fanelli-aurell}
we establish the density profile of a collapsing
cluster. In Section \ref{s:2clust} we study the
dynamics of two massive clusters, and in Section~\ref{s:con} we sum up
and discuss our results.


\section{From Vlasov-Poisson equations to the adhesion model}
\label{s:VP}

The evolution of collision-less dark matter in a three dimensional
expanding Universe is described by the kinetic Vlasov-Poisson equations.
Assume an inertial reference frame, and label the position with ${\bf r}$.
Then  it is customary to introduce the co-moving coordinate ${\bf x}$
by the following transformation \cite{Peebles1980}:
\begin{equation}
\label{transf}
{\bf r} = a(t) {\bf x}~,
\end{equation}
where the scale factor $a(t)$ is function of
proper world time. For an Einstein-de Sitter Universe
\begin{equation}
\label{scale}
a=\left(\frac{t}{t_0}\right)^{2/3}~,
\end{equation}
where $t_0^{-2}=6 \pi G \rho(t_0)$, $\rho(t_0)$ is the homogeneous
density at time $t_0$
\cite{Peebles1980,Weinberg1972} and $G$ is the
gravitational constant. The  Vlasov-Poisson equations then read:
\begin{equation}
\label{eq:system-JVP2}
\left\{
\begin{array}{l}
\displaystyle{\partial_t f+\frac{\bf p}{m a^2}\cdot
{\bf \nabla_x} f -{\bf \nabla_x}\psi\cdot
{\bf \nabla_p} f = 0}
\\
\\
\displaystyle{\nabla^2\psi=4 \pi G a^2 \left( \rho - \rho_b \right)}~,
\end{array}
\right.
\end{equation}
where ${\bf p}$ is the variable conjugated to ${\bf x}$;
$f({\bf x},{\bf p},t)$ is the distribution
function in the six-dimensional phase space $({\bf x},{\bf p})$;
$\psi$ is the gravitational potential and $\rho_b$ is
the mean mass density.
The particle density $\rho({\bf x},t)$ and velocities ${\bf u}({\bf x},t)$
are given, in term of $f({\bf x},{\bf p},t)$, as:
\begin{equation}
\label{rho0}
\rho({\bf x},t)=\frac{m}{a^3} \int f({\bf x}, {\bf p},t) d{\bf p}~,
\end{equation}
\begin{equation}
\label{u}
\rho({\bf x},t) {\bf u}({\bf x},t) =
\frac{1}{a^4} \int {\bf p} f({\bf x},{\bf p},t) d{\bf p}.
\end{equation}
It is well known, see e.g.~\cite{Vergassola}, that (\ref{eq:system-JVP2}) admit
special solutions of the form
\begin{equation}
\label{singlespeed}
f({\mathbf x},{\mathbf p},t) = \frac{a^3 \rho({\mathbf x},t)}{m} \delta^d ({\mathbf p}-m a
{\mathbf u}({\mathbf x},t))~,
\end{equation}
where $d$ is the dimension of space and $\delta^d(.)$ the $d$-dimensional
delta function. We will refer to this class as to single-speed solutions,
because to each given (${\mathbf x},t$) corresponds a well defined velocity
${\mathbf u}$. Assuming (\ref{singlespeed}), a closed system on the hydrodynamic
level can be derived using
(\ref{rho0}) and (\ref{u}):
\begin{equation}
\label{s:system}
\left\{
\begin{array}{l}
\displaystyle{\partial_t\rho+3\frac{\dot{a}}{a}\rho +
\frac{1}{a}{\mathbf \nabla} \cdot  ({\rho} {\mathbf u})=0}
\\
\\
\displaystyle{\partial_t {\mathbf u}+
\frac{\dot{a}}{a}{\mathbf u}+
\frac{1}{a}({\mathbf u} \cdot {\mathbf \nabla})
{\mathbf u} = {\mathbf g}}
\\
\\
\displaystyle{{\mathbf \nabla} \cdot {\mathbf g}=-4 \pi G a (\rho-\rho_b)~,}
\end{array}
\right.
\end{equation}
where we have introduced ${\mathbf g} = - {\mathbf \nabla} \psi/a$ such that
${\mathbf \nabla} \times {\mathbf g} =0$. It should be stressed that system
(\ref{s:system}) is valid as long as the distribution function
$f({\mathbf x},{\mathbf p}, t)$ is in the form (\ref{singlespeed}).
Beyond the time of caustic formation, when fast particles cross slow
ones, the solution becomes multi-stream.
Hence, the pressure-less and dissipation-less hydrodynamical equations of
(\ref{s:system}) are incomplete.
\\\\
An ansatz that permits further progress is the so-called 
\textit{condition of parallelism} which requires that
the peculiar velocity is a potential field, that remains
parallel to the gravitational peculiar acceleration field
\cite{Peebles1980,Vergassola,Buchert2}:
\begin{equation}
\label{parallel}
{\mathbf g}=F(t){\mathbf u}~
\end{equation}
The important assumption, in the present discussion, is that 
the proportionality $F(t)$ is space-independent.
That can only strictly be true in the linear regime, where
\begin{equation}
\label{parall}
F(t)=4 \pi G \rho_b b/\dot{b}~,
\end{equation}
and $b$ is the amplitude of the
perturbation.
A natural further assumption is that only
the linearly growing mode is excited~\cite{Peebles1980,Buchert2}, 
which gives $b\sim t^{\frac{2}{3}}$ in our case.
Defining the new velocity field
${\mathbf v}={\mathbf u}/(a \dot{b})$, system (\ref{s:system}) 
reduces to free motion in Eulerian coordinates
\begin{equation}
\label{burgers}
\partial_b {\mathbf v} + ({\mathbf v} \cdot {\mathbf \nabla}) {\mathbf v}=0~,
\end{equation}
This is Zeldovich' pancake model in the original formulation~\cite{Zeldovich}. 
It is also to referred as the Zeldovich approximation, because
the contribution of the decaying mode of the density field 
has been neglected. 
The ansatz that the condition of parallelism holds also in the
multi-stream region could be approximatively true, but is still
a bold assumption to make.
\\\\
After caustic formation one may think that the resulting
change in the gravitational force could be
modeled by an effective diffusive term.
In the adhesion model, \cite{Gurbatov,libro}, 
one hence introduces a term of the form
$\nu \nabla^2 {\mathbf v}$ in the right hand side of the equation
(\ref{burgers}): 
\begin{equation}
\label{s:adhesion}
\left\{
\begin{array}{l}
\displaystyle{\partial_b {\mathbf v} + ({\mathbf v} \cdot {\mathbf \nabla}) {\mathbf v}
= \nu \nabla^2 {\mathbf v} }
\\
\\
\displaystyle{{\mathbf v}= - {\mathbf \nabla} \tilde{\psi}}
\\
\\
\displaystyle{\partial_b \rho + {\mathbf \nabla} \cdot (\rho {\mathbf v})=0~,}
\end{array}
\right.
\end{equation}
where $\tilde{\psi} = \psi/(\dot{b} F(t) )$. 
In order for the diffusion term to have a smoothing effect
only in those regions where the particles crossing takes place, the 
phenomenological viscosity parameter
$\nu$ should be small. 
\\\\
Although numerical experiments suggest qualitative agreement,
the exact relationship between (\ref{eq:system-JVP2}) and (\ref{s:adhesion})
is still an open problem. In a recent paper \cite{fanelli-aurell} a
derivation of (\ref{s:adhesion}) was outlined
in one dimension.
Comparisons with numerical results suggested however 
that at least the diffusion term should
be modified, to give instead 
transport equation in the class recently studied
by Buchert and co-workers~\cite{Buchert1,Buchert2}.
We will here study the system (\ref{eq:system-JVP2}) directly,
and establish the inner structure of a mass agglomeration without
the ansatz of parallelism.


\section{The Q model, and its relation to the 
Zeldovich approximation and the adhesion model} 
\label{Q}

We will consider the evolution of the one-dimensional perturbation
in the expanding three dimension Universe, using the discrete
approximation of initial condition. Let us assume, that in some
interval $L$ along the axis $x$ all the matter is concentrated in
the $N$ sheets, which we will from now on
call particles.
\\\\
The Newtonian equations of motion for such $N$ particles, interacting via gravity,
follow from the Lagrangian \cite{Peebles1980}:
\begin{equation}
\label{background_lagrangian}
{\mathcal{L}} = \sum_i \frac{1}{2} m_i \dot{{r_i}}^2 -
m_i \phi ({r_i},t)~,
\end{equation}
where $\nabla_r^2\phi = 4\pi G\rho$. In the point particle picture the
density profile reads:
\begin{equation}
\label{rho}
\rho(x_i,t)=\sum_{x_j} m_j a^{-3} \delta(x_i-x_j)~,
\end{equation}
where $x_i$ is the co-moving coordinate of the $i$'th particle, in the
direction of which the density and velocities vary.
\\\\
Expressing (\ref{background_lagrangian}) as a function of the proper
coordinate, $x_i$, and assuming (\ref{rho}), the equation of motion of the
$i$'th  particle reads:
\begin{equation}
\label{euler1}
\frac{d^2 x_i}{dt^2} + 2 \frac{\dot{a}}{a} \frac{d x_i}{dt}
-4 \pi G  \rho_b(t) x_i = a^{-3} E_{grav}(x_i,t)~,
\end{equation}
where $\rho_b(t)$ is the mean mass density at time $t$ and
\begin{equation}
\label{selfgrav}
E_{grav} (x_i,t) = - 2 \pi G  \sum_j m_j \hbox{sign} (x_i-x_j)~.
\end{equation}
In an Einstein-de Sitter Universe we can make a nonlinear
change of variable, $\tau = t_0 \log t/t_0$, such that
(\ref{euler1}) takes the form:
\begin{equation}
\label{euler3-Q}
\frac{d^2 x_i}{d \tau^2} +  \frac{1}{3t_0} \frac{d x_i}{d \tau}
-  \frac{2}{3t_0^2} x_i =  E_{grav} (x_i,\tau) \qquad\hbox{Q model}~,
\end{equation}
where $t_0^{-2} = 6 \pi G \rho(t_0)$.
This is the representation of the
Vlasov-Poisson equations which we call the Quintic (Q) model.
\\\\
The interest of this formulation is that, as for the
classical static self-gravitating systems in one dimension,
$E_{grav}$ is a Lagrangian invariant, proportional to the net mass
difference to the right and to the left of a  given particle, at a given
time. Hence, in between collisions, (\ref{euler3-Q}) has
an explicit solution~\cite{Qmodel}:
\begin{equation}
\label{global}
x_i(\tau) = c^i_1 \exp(\frac{2(\tau-\tau^n)}{3t_0}) + c^i_2 \exp(-\frac{(\tau-\tau^n)}{t_0})+ K_i^n~,
\end{equation}
where $K_i^n=-(3t_0^2 / 2) E_{grav} (x_i,\tau)$, is constant.
The coefficients $c^i_1$ and $c^i_2$ are determined by $x_i^n=x_i(\tau^n)$
and $w_i^n=\dot{x_i}(\tau^n)$, i.e. by the states of the particle at the
time of the last crossing, and read:
\begin{equation}
\label{s:coeff0}
\left\{
\begin{array}{l}
\displaystyle{c^{i}_1 =  \frac{3}{5} \left[ x_i^n + t_0 w_i^n  - K_i\right]}
\\
\\
\displaystyle{c^{i}_2 = \frac{2}{5} \left[ x_i^n - \frac{3}{2} t_0 w_i^n  -
K_i \right]  }~.
\end{array}
\right.
\end{equation}
The form of equation (\ref{global}) suggests introducing an
auxiliary variable $z=\exp( (\tau-\tau^n) / 3t_0 )$. The crossing times
between neighboring particles (i.e. $i$,$i+1$) can, hence, be computed by
solving numerically the following \textit{quintic} equation:
\begin{equation}
\label{quint}
f(z) = A_{i,i+1}^n z^5 - B_{i,i+1}^n z^{3} + C_{i,i+1}^n = 0~,
\end{equation}
where:
\begin{equation}
\label{s:coeff}
\left\{
\begin{array}{l}
\displaystyle{A_{i,i+1}^n =  \frac{3}{5} \left[
\Delta x_{i}^n + t_0 \Delta w_{i}^n  - (K_{i+1}^n-K_i^n) \right]}
\\
\\
\\
\displaystyle{B_{i,i+1}^n =  -(K_{i+1}^n-K_i^n) = const}
\\
\\
\displaystyle{C_{i,i+1}^n =\frac{2}{5} \left[
\Delta x_{i}^n -\frac{3}{2} t_0 \Delta w_{i}^n  - (K_{i+1}^n-K_i^n) \right]}~,
\end{array}
\right.
\end{equation}
and $\Delta x_{i}^n=x_{i+1}^n-x_i^n$,  $\Delta w_{i}^n=w_{i+1}^n-w_i^n$.
Thus, the evolution of the system is recovered by using a version
of the event-driven scheme discussed in~\cite{al} and~\cite{japan}.
Details of the numerical implementation are given in
\cite{fanelli-aurell}.
\\\\
Let as assume that in the interval $L$ all the particles have equal
mass $m_j=L \rho_0/N$. Then, the gravitational force in equation (\ref{euler3-Q})
transforms in
\begin{equation}
\label{selfgravN}
E_{grav} (x_i,\tau) = - \frac{L}{3Nt_0^2}  \sum_j  \hbox{sign} (x_i-x_j)=\frac{L}{3Nt_0^2}N_i ~.
\end{equation}
where $N_i=N_{i,right}-N_{i,left}$ is the difference between the number of
particles to the right and to the left of a  given particle. At the
initial time $\tau=\tau^0=0$, the positions of all the particles
are ordered ($x_{i+1}> x_{i}$), and
we have $N_i=N-(2i-1)$. With $K_i^0$ in the solution (\ref{global})
constant we get
\begin{equation}
\label{constK}
K_i^0 = -\frac{L}{2N}N_i^0= -\frac{L}{2}+ \frac{L}{N}(i- \frac{1}{2}) ~.
\end{equation}
If we now pick the specific initial conditions such that all the
$c_2^i$'s in (\ref{global}) are zero,
we recover, in the discrete setting, the special case of
correlated velocity and density perturbations, such that
only the linearly increasing mode is excited. Note that this separation
is valid until the first particle crossing, i.e. well into the 
non-linear regime. In this time interval we have
\begin{equation}
\label{globalN}
x_i(\tau) = \frac{3}{2}w_i^0 t_0 \exp(\frac{2\tau}{3t_0}) - \frac{L}{2}+ \frac{L}{N}(i- \frac{1}{2})~,
\end{equation}
Hence, defining the new ``time'' $b=b_0\exp( 2 \tau / 3t_0)$
(with density dimension),
the solution of (\ref{globalN}) reduces to free motion of the particles.
In other words, we have re-derived Zeldovich' approximation
in the framework of the Q model.
In the Zeldovich approximation, this solution is assumed to hold for all time,
even when the solution becomes multi-stream, while in the adhesion model we
have merging of the particles after their crossing. The real system must
necessarily lie somewhere in between.


\section{A two particles system: an iterative map}
\label{s:2party}

\begin{figure}[ht!]
\begin{center}
\psfig{figure=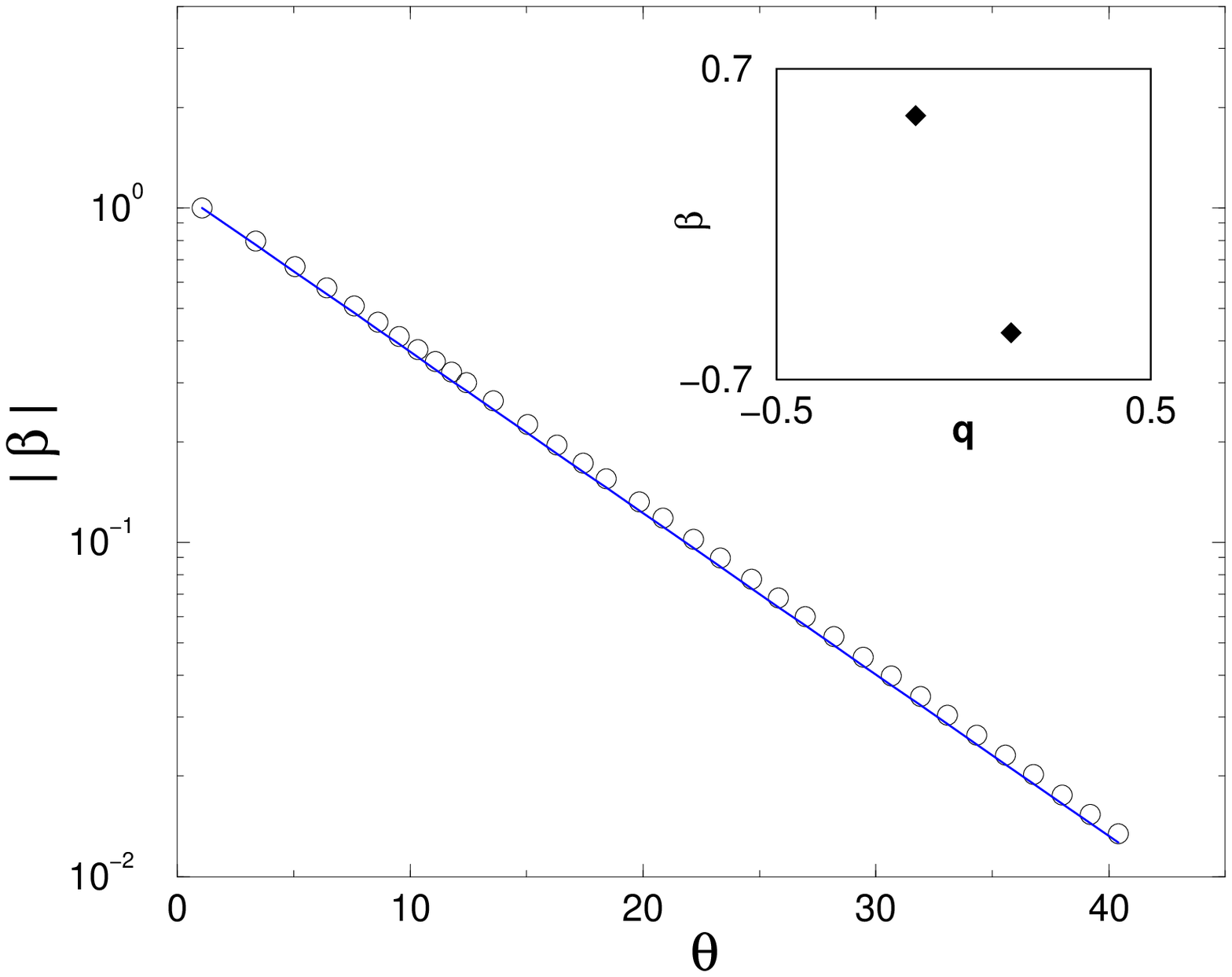,height=10truecm,width=10truecm}
\caption{\label{fig1} The absolute value of the velocity $\beta_{cross}$ is
plotted as function of rescaled time $\theta$, in scale lin-log. Circles refer to the
simulations, while the solid line is the theoretical prediction (\ref{beta_osc}). The initial
condition is displayed in the small inset.}
\end{center}
\end{figure}

\begin{figure}[ht!]
\begin{center}
\psfig{figure=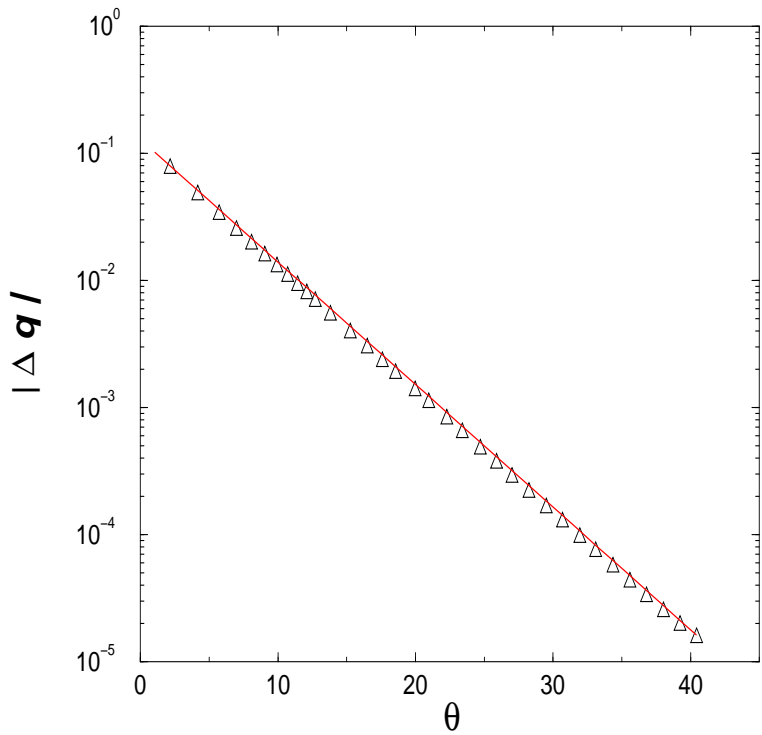,height=10truecm,width=10truecm}
\caption{\em \label{fig2} The inter-particle distance $|\Delta q_{max}|$ is represented as function of rescaled
time $\theta$, in scale lin-log. Triangles refer to the simulation, while the solid line is the
theoretical prediction (\ref{deltaq_osc}).}
\end{center}
\end{figure}

Consider two particles of equal mass $m$ interacting by the Q dynamics,
confined in a box of size $L$. 
We choose units of length and mass such that $L=1$ and $m=1/2$.
To streamline the following, we first
first introduce the dimensionless variables:
\begin{eqnarray}
\label{dimlessvar}
\theta &=& \frac{\tau}{t_0}~,\\
\
q_i(\theta) &=& \frac{x_i}{L}~,\\
\
\beta_i(\theta) &=& \frac {6 {\dot x_i}t_0 }{L}.
\end{eqnarray}
Thus, before the time of first crossing,  equation (\ref{global}) takes the form:
\begin{equation}
\label{global1}
q_i(\theta) = c^i_1 \exp(\frac{2\theta}{3}) + c^i_2 \exp(-\theta) \mp \frac{1}{4}~.
\end{equation}
where $c_1^i$ and $c_2^i$ are given by (\ref{s:coeff0}).
We select a special class of initial condition: the velocities of the particles are arbitrarily
assigned, while the positions are determined by setting $c^i_2$ to zero (\ref{globalN}).
This simply means that there is no decaying mode initially, as discussed in the
end of the previous section, and we can therefore immediately
integrate the equations of motion of every particle up to its first crossing.
In addition we chose a system of reference such that the center of mass is
at rest at the origin. That means that the initial
velocities are $w_1^0 = -w_2^0 > 0$. 
For simplicity we will suppress in the following the index $i$ and consider the
evolution of the leftmost of the two particles.
\\\\
The time of first crossing, $\theta_{cross}$, is then deduced 
from equation (\ref{global1})
and reads 
\begin{equation}
\label{teta_col}
\theta_{cross}=-\frac{3}{2} \ln{\beta(0)}.
\end{equation}
The rescaled particle's velocity is  $\beta_{cross,0}=\beta(\theta_{cross})= 1$.
Remark that $\theta_{cross}$ is solely determined by the initial particle
velocity, while $\beta_{cross,0}$ is a dimension-less constant, not
dependent on the states of the particles at $\theta=0$. 
This means that while
the time at which the particles cross ($\theta_{cross}$) depends
on the initial velocity, their further evolution does not.
On the other hand, after crossing the 
initial balance between positions and velocities no longer holds,
and the condition $c_2^i=0$ is no longer true.
From that time on
both the growing and decaying modes in (\ref{global1}) have to be
considered in the analysis.
\\\\
For times larger than $\theta_{cross}$, the separation between
the particles increases. Conversely,
as an effect of the gravitational attraction, their velocities are progressively reduced.
Hence, a turning point is reached where the elongation is maximal and the velocities are
zero. The inversion time, $\theta_{turn}$, follows by differentiating equation (\ref{global})
with respect to time, and imposing the condition $\dot{q}(\theta_{turn})=0$. Thus:
\begin{equation}
\label{teta_max_1}
\theta_{turn} = \theta_{cross} + \tilde{\theta}
\end{equation}
where $\tilde{\theta}=\frac{3}{5} \ln{6}$.
The corresponding maximum elongation, $q_{max,0}=q(\theta_{turn})$, reads:
\begin{equation}
\label{q_max_1}
q_{max,0} = \frac{1}{4} - \frac{1}{5} 6^{-\frac{3}{5}} -\frac{1}{20} 6^{\frac{2}{5}} = \alpha \simeq 0.08.
\end{equation}
It does not depend on the initial velocity.
\\\\
Note that the particle inversion occurs before the {\it initial position}
(i.e. $|q_i(0)|= \frac{1}{4}(1-\beta(0)$)
is recovered: this is the effect of the friction-like term in
equation (\ref{euler3-Q}), which is responsible for the progressive 
localization of the
system around the center of mass. An analogous damping is shown to occur for the
particles velocities. In fact, for $\theta>\theta_{turn}$, the particles move inward
approaching the time of successive crossing. Due to the inner symmetry, the latter will take place in
the origin. The particle rescaled velocity  corresponding to the
{\it first return} reads:
\begin{equation}
\label{beta_ret}
\beta_{ret}=\beta_{cross,1}=-0.83\beta_{cross,0}=-0.83.
\end{equation}
By a simple generalization of the previous discussion, 
it can be shown that, for larger times, 
the particle executes damped oscillations.
The particles move back and forth, displaying a progressive reduction
of their mutual distance: the crossings become more frequent as time is increased.
Take an index $n$ to label particle crossings in the system. The variable
$\theta_n$ represents the time elapsed between the 
$n-1$'th and $n$'th crossings, and
$\beta_{cross,n}$ is the dimensionless velocity of the particle
at the $n$'th crossing.
Then we make the further assumptions of smallness:
\begin{eqnarray}
\label{iter_appr}
|\beta_{cross,n}| \ll 1.
\end{eqnarray}
For large enough $n$ these are reasonable approximation.
Further, we assume (\ref{iter_appr}) to hold even for an early stage of
the evolution: we will return on this point, and provide numerical
support to the validity of this ansatz. Within these approximation the
dynamics of (\ref{global}) is reduced to the
following simple map:

\begin{equation}
\label{map}
\begin{array}{lcl}
\beta_{cross,n+1} &=& -\beta_{cross,n} \left(1 - \frac{4}{3}\beta_{cross,n}\right)\\
\theta_{n+1} &=& 2 |\beta_{cross,n}| \\
\Delta q_{max,n+1} &=& \frac{1}{12} \beta_{cross,n}^2 
\end{array}
\end{equation}
where $\Delta q_{max,n}$ stands for the maximal elongation reached by the particles
after $n$ encounters. 
It is interesting that system (\ref{map}) is actually equivalent
to the motion of an impurity advected by an unsteady Burgers flow with a particular
numerical relation between frictional and inertial forces, see \cite{GuMaiTaMo}.
\\\\
Consider then the finite difference formula:
\begin{equation}
\label{diff}
\frac{\Delta |\beta_{cross}|}{\Delta\theta}=\frac{|\beta_{cross,n+1}|-|\beta_{cross,n}|}{\theta_{n+1}},
\end{equation}
and rewrite the right hand-side using the first two relations of (\ref{map}).
Approximating the finite differences with differentials and
integrating, using as initial condition for 
$|\beta_{cross}(\theta=\theta_{cross})|$ the velocity of at 
first crossing $|\beta_{cross,0}|$ and for 
$|\Delta q_{max}(\theta=\theta_{turn})|$ the first maximum 
elongation (\ref{q_max_1}), one ends up with:

\begin{equation}
\label{beta_osc}
|\beta_{cross}(\theta)| = |\beta_{cross,0}| \exp \left( -\frac{1}{9}(\theta-\theta_{cross}) \right),\,\,\,
\beta_{cross,0}=1,
\end{equation}

\begin{equation}
\label{deltaq_osc}
|\Delta q_{max,0}(\theta)| = 2 q_{max} \exp \left( -\frac{2}{9}(\theta-\theta_{turn}) \right),\,\,\,\,
q_{max,0}= \alpha.
\end{equation}
Here $|\beta_{cross}(\theta)|$ is the dimension-less velocity of the particles
at the moment of their crossing and $\Delta q_{max}(\theta)$ the maximum separation between the two particles, at rescaled time
$\theta$. 
Both the velocity and the inter-particle distance 
(and also the time elapsed between two successive crossing  $\theta_{n}(\theta) \sim |\beta_{cross}(\theta)|)$ 
decrease exponentially, as function
of time $\theta$, or, equivalently, a power-law decays is displayed in term of cosmological
time $t$: $|\beta_{cross}(t)| \sim t^{-1/9}$ and $|\Delta q_{max}(t)| \sim t^{-2/9}$.
\\\\
Let us turn to numerical simulations to test the reliability of our results. The initial
condition is displayed in the small inset of Figure \ref{fig1}: here $c_2=0$, thus
before the time of first crossing, only the contribution of the growing mode has to be considered.
In the main plot of Figure \ref{fig1} the absolute value of the particle velocity, $\beta_{cross}$, is represented
as function of rescaled time  $\theta$, in scale lin-log. The solid line refers to the analytical
prediction (\ref{beta_osc}).
In Figure \ref{fig2} the particle maximum separation $\Delta q_{max}(\theta)$ is plotted versus $\theta$,
in linear-logarithmic scale (\ref{deltaq_osc}). The solid line shows the theoretical result
(\ref{deltaq_osc}). In both case good agreement is displayed, even in the initial stage of the
dynamics, thus confirming a posteriori the validity of our assumptions.


\section{Density profile: Q model vs. adhesion model}
\label{s:spatialdens}

The objective of this section is to investigate the density profile
of a collapsing cluster in the Q model, and compare that to the
adhesion model. We show that collapse in the Q model,
which we recall is nothing but one-dimensional self-gravitating dynamics
on the background of an Ein\-stein-de~Sitter Universe, is
less pronounced. We derive a general expression for 
the density profile. 
To this end, we begin by analysing
the motion a particle in an external halo, moving in the
potential from a much more massive and very
localized agglomeration. Then,
we reconstruct the whole density distribution by combining together
the contribution from each particle. 
These theoretical results are then compared 
first with numerical simulations of the Q model,
and then with predictions based on 
the Zeldovich approximation and the adhesion model.
\\\\
Consider a system of $N$ particles of equal mass, confined in box of
finite size $L=1$, symmetric with respect to a reference point located
at the origin.
First, assume $N-2l$ particles to form an inner bulk of width $\Delta$,
localized in the center of the box, with $l$ much smaller than $N$,
and $\Delta$ much smaller than $L$.
The remaining $2l$ particles are assumed uniformly distributed in
the outer regions at the borders of the box. 
\\\\
To first approximation we can neglect the fine structure of the cluster,
and replace it with an heavy, structure-less, {\it macro-particle}, 
carrying the same amount of mass. In addition,  the gravitational interaction 
between the particles in the halo is also neglected.
As a consequence, these particles are then not affected by
the discreteness of the inner distribution: 
a sudden change in the acceleration is experienced when the origin 
is crossed.
\\\\
As in the previous section, we set initial conditions such that only
the growing mode is active, before the first particles crossing occurs,
and use the same dimension-less variable.
Consider the motion of particle $i$, situated to the left of the
agglomeration and define:

\begin{equation}
\label{gamma}
\gamma_i = \frac{1}{2} - \frac{1}{N}(i- \frac{1}{2})~.
\end{equation}
The positive parameter $\gamma_i$ represents the absolute value of the initial 
position of the $i$'th particle ($w_i^0=0$ in (\ref{globalN})).
In analogy with (\ref{teta_col})-(\ref{beta_ret}), one finds:

\begin{equation}
\label{itheta}
\theta_{i,cross}=-\frac{3}{2} \ln \left( \frac{ \beta_i}{4 \gamma_i} \right)~,
\end{equation}

\begin{equation}
\label{ibeta}
\beta_{i,cross}=\beta_i(\theta_{cross})= 4\gamma_i~,
\end{equation}

\begin{equation}
\label{iturn}
\theta_{i,turn}=\theta_{i,cross} + \tilde{\theta}~=\theta_{i,cross} + \frac{3}{5} \ln{6}~,
\end{equation}

\begin{equation}
\label{iqmax}
q_{i,max,0}= 4 \alpha \gamma_i~.
\end{equation}
In this approximation,  the $i$'th particle crosses for the first time the 
origin with rescaled velocity $\beta_{i,cross,0}$. Note that the latter is 
solely determined by the particle initial position. The
corresponding maximum elongation $q_{i,max,0}$
is $4 \alpha \simeq 0.32 $ times smaller then the unperturbed  
distance $\gamma_i$. 
\\\\
By repeating the same procedure as in Section \ref{s:2party},
one ends up with the following relations for, respectively,
$|\beta_{i,cross}|$ and $|\Delta q_{i,max}|$:

\begin{equation}
\label{beta_osc_clust}
|\beta_{i,cross}(\theta)| = 4\gamma_i \exp \left( -\frac{1}{9} (\theta-\theta_{i,cross}) \right),
\end{equation}

\begin{equation}
\label{deltaq_osc_clust}
|\Delta q_{i,max}(\theta)| =4 \alpha \gamma_i~ \exp \left( -\frac{2}{9}(\theta-\theta_{i,turn}) \right).
\end{equation}

\begin{figure}[ht!]
\begin{center}
\psfig{figure=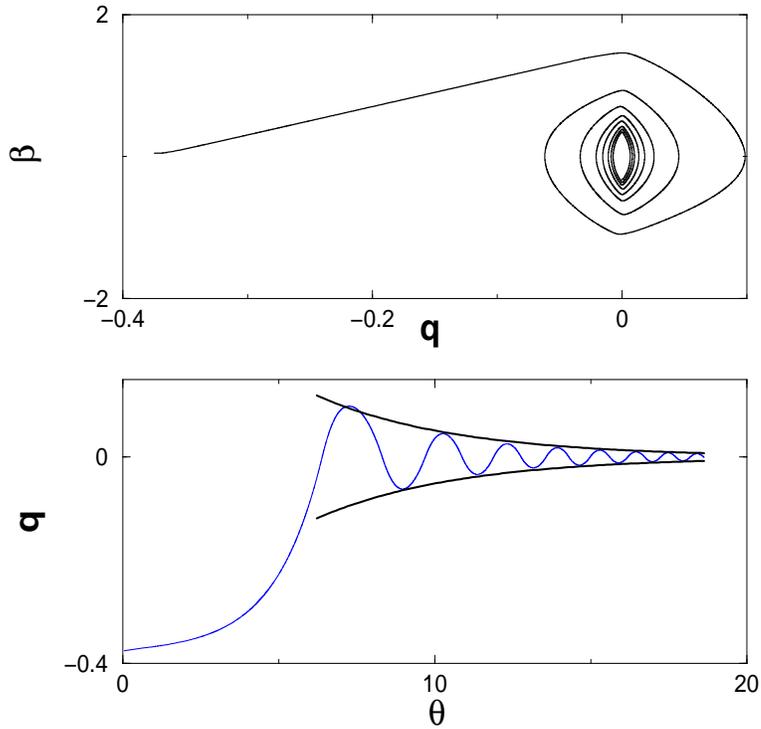,height=10truecm,width=10truecm}
\caption{\label{fig4} \em Upper panel: phase space portrait of a single particle.
Lower panel: position $q$ vs. rescaled time $\theta$. The thin solid line refers to the simulation while the thick curve represent the theoretical prediction (\ref{deltaq_osc_clust}). No parameter needs to be adjusted by numerical
fitting. Here $N=1024$, $N-2\ell=768$, $\Delta=0.15$.}
\end{center}
\end{figure}

To validate our analysis we performed numerical simulations and studied the single particle
behavior. In our experiments we considered the following initial condition:
a finite, but large, fraction of 
all the particles, say $N-2\ell$, is uniformly distributed in a narrow region,
say $\Delta$, centered around the
reference origin. The remaining particles, $\ell$ of them, are symmetrically placed
in the two lateral regions, with
spatially uniform distribution.
As a consequence, the inter-particle distance in the central bulk 
is smaller than in
the external halos.
In the limit when $\Delta \rightarrow 0$ and $\ell=1$,
we reach the conditions assumed in the preceding derivation.
Velocities are given as a smooth function of positions. 
\\\\
In the upper panel of Figure \ref{fig4} a phase space portrait of a single particle is
represented: the damping, both in $q$ and $\beta$, is clearly displayed. In the lower panel, the rescaled
particle position $q$ is plotted vs. time $\theta$: the thin solid line refers to the simulation, while
the thick curve is plotted from relation (\ref{deltaq_osc_clust}).
Numerical simulations show good agreement with analytical predictions 
(\ref{beta_osc_clust}) and (\ref {deltaq_osc_clust}).
\\\\
The previous results can also be extended to the case when particles are
initially uniformly distributed in the finite size box. Given particle
j, we neglect the interaction with the external masses 
(i.e. $|q_j|>|\gamma_i|$) and mimic the attraction of the inner particles 
by considering the forces exerted by a single massive object, localized in
the origin. In this approximation, according to the lines of the preceding
discussion,  the behaviors of the $|\beta_{i,cross}(\theta)|$ and $|\Delta
q_{i,max}(\theta)|$ are also described by  equations
(\ref{beta_osc_clust}) and (\ref {deltaq_osc_clust}). Despite the drastic
assumptions involved, numerical simulations show relatively good agreement
with theoretical predictions, provided the velocity distribution are
smooth enough.

The goal of the remaining discussion is to derive an analytical estimate of 
the density profile by making use of these results. In the continuum limit
($N \rightarrow \infty $), the density $\rho(q,\theta)$ reads:

\begin{equation}
\label{J}
\rho(q,\theta)=\rho_0 /|\Delta q/\Delta \gamma| ,
\end{equation}
where we introduced the Jacobian of the transformation from Lagrangian 
to Eulerian coordinates. The density field has a rather complex structure
including singularities $\rho(q,\theta) \sim (q-q_m)^{-1/2}$, where 
$q_m$ stands for the particle turning point.
A rough characteristics of the cluster size, $q_{s}$, can 
be deduced from equation (\ref{iqmax}). In fact, before the 
time the outmost particle 
reaches, as the last one, the massive bulk, the size of the 
agglomeration, $q_{s}$,
is given by:

\begin{equation}
\label{qwthete}
q_{s}(\theta)=4 \alpha \gamma_{i*}(\theta)~,
\end{equation}
where $-\gamma_{i*}(\theta)$ is the initial  coordinate 
of the particle which falls into the origin at the time
$\theta -\tilde{\theta}$. From (\ref{itheta}), (\ref{iturn}) it follows:

\begin{equation}
\label{gamma*}
\gamma_{i*} (\theta) = \frac{ \beta_{i*}}{4}\exp \left( \frac{2}{3} (\theta -\tilde{\theta}) \right)=
 \frac{ \beta_{i*}}{6^{2/5} 4}\exp \left( \frac{2}{3} \theta \right)
\end{equation}
When the continuous limit is recovered, 
$\beta_{i*}=\beta(-\gamma_{*})$,  $\gamma_{*}$ being the Lagrangian 
coordinate of the most periferic particle of the cluster, at time  $\theta$. 
Then assume that $\beta(-\gamma) \rightarrow 0$, as the edge of the box is 
approached ($\gamma \rightarrow 1/2$). Then from (\ref{gamma*}),  
$\gamma_*$ tends to $1/2$, at large times,  and 
consequently, from (\ref{qwthete}),   
$q_{s}(\theta)\rightarrow 2\alpha \simeq 0.16$.
This observation on is in good agreement  with the results of numerical 
simulations presented in a in previous work \cite{fanelli-aurell}.
\\\\
Inside the pancake a multistream flow has developed: the density 
at coordinate $q$ follows 
by summing the contribution of each stream.  
Consider the {\it mean} density distribution assuming the length scale 
$\Delta q$ much larger than the distance between two successive peaks i.e.
larger than $|q_{m+1}-q_m|$. The sum can be 
then approximated by an integral over 
the Lagrangian coordinate $\gamma$. 
By introducing  the  density distribution 
function $f_{\rho}(\gamma,\theta)$
associated to the Lagrangian coordinate $\gamma$, 
we can write:

\begin{equation}
\label{ro_def}
\rho(q,\theta) = \int_{\gamma_{min}}^{\gamma_{max}}f_{\rho}(\gamma,\theta) d \gamma.
\end{equation}
Assume that all the particles belonging to a small Lagrangian interval, 
$\Delta \gamma$, are uniformly distributed inside the corresponding Eulerian 
segment $\Delta q_{max}(\theta,\gamma)$ (\ref{deltaq_osc_clust}):
\begin{equation}
\label{deltaq_max}
\Delta q_{max}(\theta,\gamma)=|\Delta q_{i,max}(\theta)| 
=\frac{\chi\gamma^{\frac{4}{3}}}{\beta^{\frac{1}{3}}(\gamma)}
\exp \left( -\frac{2}{9}\theta\right),
\,\,\, \chi=4^{4/3} 6^{2/15} \alpha 
\end{equation}
and consider the mass conservation:
\begin{equation}
\label{masscons}
\rho_0 \Delta \gamma = \left(f(\gamma,\theta) \Delta \gamma \right) \Delta q_{max}(\theta,\gamma),
\end{equation}
The interval  $[\gamma_{min}(q,\theta), \gamma_{max}(\theta)]$ selects
the particles that contributes to the density in $q$, at time $\theta$.
In particular they should have reached the origin, thus:

 \begin{equation}
\label{gamma_max}
\frac{\beta(\gamma_{max})}{4}\exp \left( \frac{2}{3} \theta\right) = \gamma_{max}.
\end{equation}
On the other hand the amplitudes of the particles' oscillations, 
$\Delta q_{max}(\theta,\gamma)$ have to be larger than $q$, thus 
$\gamma_{min}$ is found by solving the following implicit equation.

\begin{equation}
\label{gamma_min}
q = \chi \exp \left(-\frac{2\theta}{9}\right) \frac{\gamma_{min}^{\frac{4}{3}}}{\beta^{\frac{1}{3}}(\gamma_{min})}
\end{equation}
Finally, the density reads:

\begin{equation}
\label{ro_general}
\rho(q,\theta) = \rho_0 \int_{\gamma_{min}}^{\gamma_{max}} \frac{d \gamma}{\Delta q_{max}(\theta,\gamma)}=
\frac{\rho_0}{\chi} \exp \left( \frac{2}{9}\theta \right) \int^{\gamma_{max}}_{\gamma_{min}} \frac{\beta^{\frac{1}{3}}
(\gamma)}{\gamma^{\frac{4}{3}}}  d \gamma,
\end{equation}
and $\rho(q,\theta)=0$ when: 
\begin{equation}
\label{widi}
|q|>q_w(\theta)=\Delta q_{max}(\theta,\gamma_{max}).
\end{equation}
Here $q_w(\theta)$ is the width of the mean density distribution. 
The density profile (\ref{ro_general}) depends on the initial particle velocity, namely
$\beta(\gamma)$. In the following, we will solve analytically the integral in (\ref{ro_general})
for special class of initial conditions, and compare with numerics.
In addition, we will discuss the analogous case in the adhesion model.


\subsection{Step Profile of Initial Velocities}
\label{sb:stepprofile}

Assume the initial velocities profile to be given by:
\begin{equation}
\label{step}
\beta(q) = -\beta_0 \hbox{sign}(q),
\end{equation}
where $\beta_0$ is a constant coefficient and, without loss of generality, we
focus on the region $q>0$.
\\\\
First consider the evolution of the density in the Zeldovich
approximation, with initial positions as in equation (\ref {globalN}).
Since the velocity profile at $\theta=0$ is a step function, the particles
with initial negative (resp. positive) spatial coordinate will move as a
whole towards the origin, keeping their mutual distances constant.
Thus  before the
last left particle reaches the origin, a two-stream flow develops over a finite
region of width $2 |q_{w, zeld}|$, where
$|q_{w, zeld}|=\beta_0 \exp(2\theta/3)/4$. Note that the the density in
each flow is equal to the initial density  $\rho_0$.
\\\\
Then follow the development under the Q dynamics
later in time. By inserting (\ref{step}) in 
(\ref{ro_general}) and performing
the integral, one obtains:
\begin{equation}  
\label{densstep}
\rho(q,\theta)|_{|q|<q_w} = \rho_0 \frac{3}{2 q_p} \left[\left( \frac{q_p}{q}\right)^{1/4} - 
 \left(\frac{1}{2 \gamma_{max}} \right)^{1/3}   \right],
\end{equation}
where $\gamma_{max}$  is solution of equation (\ref{gamma_max}) and 
$q_p(\theta)$ is the spatial scale of inner structure 
\begin{equation}
\label{widthstep}
q_p=\frac{ \chi } {2^{\frac{4}{3}} \beta_0^{\frac{1}{3}}}  \exp \left( - \frac{2}{9} \theta \right).
\end{equation}
We recall that $q_w(\theta)$ stands for the  width of the pancake,  
see eq. (\ref{widi}). Consider also the mass function $M(q, \theta)$:

\begin{equation}
\label{massfunct}
M(q,\theta) =  \int_0^q \rho(\xi,\theta)~d\xi,
\end{equation}
Inserting eq. (\ref{densstep}) and performing the integral one ends up with:

\begin{equation}
\label{massfunct1}
M(q,\theta) = \rho_0  \left[2 \left (\frac{q}{
  q_p}\right)^{3/4}  -\frac{3}{2}\left(\frac{1}{2 \gamma_{max}} \right)^{1/3} 
  \left (\frac{q}{ q_p}\right)\right].
\end{equation}

In the initial stage of the evolution $\beta(\gamma_{max})= \beta_0$, 
thus, from equation (\ref{gamma_max}): 

\begin{equation}
\label{gamma_max1}
\gamma_{max}=\frac{\beta_0}{4}\exp \left( \frac{2}{3} \theta\right).
\end{equation}
The pancake's density distribution is therefore characterized by two 
length scales: the spatial scale of inner structure, $q_p(\theta)$,  which
decays exponentially, see (\ref{widthstep}), and the width of the 
agglomeration which grows exponentially according to:

\begin{equation}
\label{widi1}
q_w(\theta)=\frac{\chi \beta_0}{4^{\frac{1}{3}}}
\exp \left(\frac{2}{3}\theta\right) \simeq 0.1 \beta_0  \exp \left(\frac{2}{3}\theta\right).
\end{equation} 
It is worth stressing that during the initial stage of 
the pancake's formation, the width $q_{w}$ increases with time slower than
as predicted in the Zeldovich picture. 
From equation (\ref {densstep}) it follows that the 
density in the bulk
increases exponentially in time-like coordinate $\theta$, 
according to 
$\rho(q,\theta) \sim q^{-(1/4)}\exp(\theta/6)$. In other words, the
system tends
to display a progressively denser core, localized near the 
origin. The mass contained in the cluster is also increasing exponentially,
$m_p(\theta)=2 M(q_w,\theta)=\rho_0 \beta \exp(2\theta/3)/2$, in agreement
with the predictions of the Zeldovich model. 
The discussion above applies to the co-moving frame, 
but physical density is measured in the inertial
frame; the two are related by (\ref{transf}).
We will return to this point in Sect.~\ref{s:con}. Let us however
in brief state that the collapse in co-moving coordinates is
slower than the expansion of the Universe. The physical density
therefore on the contrary decreases at around almost all mass points, and 
in an inertial coordinate system the agglomerations spread out over time,
albeit more slowly than the expansion of the Universe as a whole. 
\\\\
Equations (\ref{massfunct1})-(\ref{widi1}) suggest
that, initially, both the mean density distribution and mean mass
function are self-similar. In particular:

\begin{equation}
\label{massfunctss}
M(q,\theta) = m_p   \left[2 \left (\frac{q}{q_w}\right)^{3/4}  -\frac{3}{2} 
  \left (\frac{q}{ q_w}\right)\right] = m_p(\theta) \overline{M}(q/q_w(\theta)).
\end{equation}

In the very late time evolution of an isolated cluster all
particles have reached the inner bulk (i.e. $\theta> - 3/2 \ln (\beta_0/2)$ ). 
The box length, $L$, is normalized to one so that, $m_p (\theta)=\rho_0 $. Then  
$\gamma_{max}=1/2$ and equations (\ref{densstep}),(\ref{massfunct}) indicate the 
a self-similar collapse: 
\begin{equation}
\label{mass-late}
M(q,\theta) = \overline{M}(q/q_p(\theta)) .
\end{equation}
Now both the density and the mass function are solely characterized by a unique 
length scale, which is both the inner length scale, $q_p(\theta)$,
and the outer length scale, $q_{w}$.
In this case $q_p$ is the coordinate of the out-most particle, and 
the density for $q>q_p$ is identically equal to zero.
\\\\
Summing up, we have shown
that for an initial condition in the form of equation 
(\ref{step}), the mean mass function $M(q,\theta)$, displays a self-similar 
collapse, in different stages of the evolution. Introducing the rescaled 
variable $x = q/q_w(\theta) $, the function $\overline{M}(q/q_w(\theta))$ takes the universal form:
 
\begin{equation}
\label{massfuncself}
\overline{M}(x)=\left[2 x^{3/4}  -\frac{3}{2} x \right] .
\end{equation}
for $x=[0,1]$, where $\overline{M}(1)$ is equal to one 
half. From (\ref{massfuncself}) it follows that the matter is mainly
concentrated around the pancake center. In particular,  one-half
of the whole mass (i.e. $2\overline{M}(x_{0.5})=1/2$) is found to lay 
in a symmetric interval defined by $x_{0.5}=0.14$, while $90 \%$ 
of it, is distributed in a segment equal to half the size of entire
cluster ($x_{0.9}=0.54$).

\begin{figure}[ht!]
\begin{center}
\psfig{figure=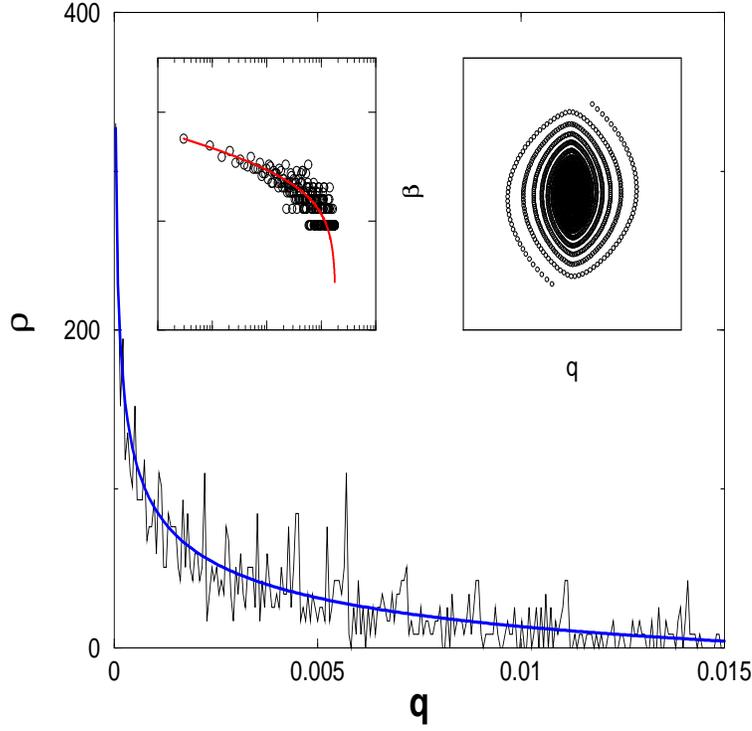,height=10truecm,width=10truecm}
\caption{\label{fig3} \em Main plot: normalized density profile $\rho$ vs. at late stage of evolution 
rescaled position $q$, starting from a
step-profile for velocities. The particles are initially uniformly distributed in space.
The thin solid line refers to the simulation, while the thick solid one represents the theoretical prediction
(\ref{densstep})($\gamma_{max}=1/2$). In this case $N=2000$, $\theta=23.27$, $\beta_0=4.9 \times 10^{-4}$. Hence, $q_p=0.0183$.
Left inset: normalized density profile $\rho$ vs. rescaled position $q$ in log-log scale. The circles refer to the
simulation. Right inset: phase space portrait at  $\theta=23.27$.}
\end{center}
\end{figure}

\begin{figure}[ht!]
\begin{center}
\psfig{figure=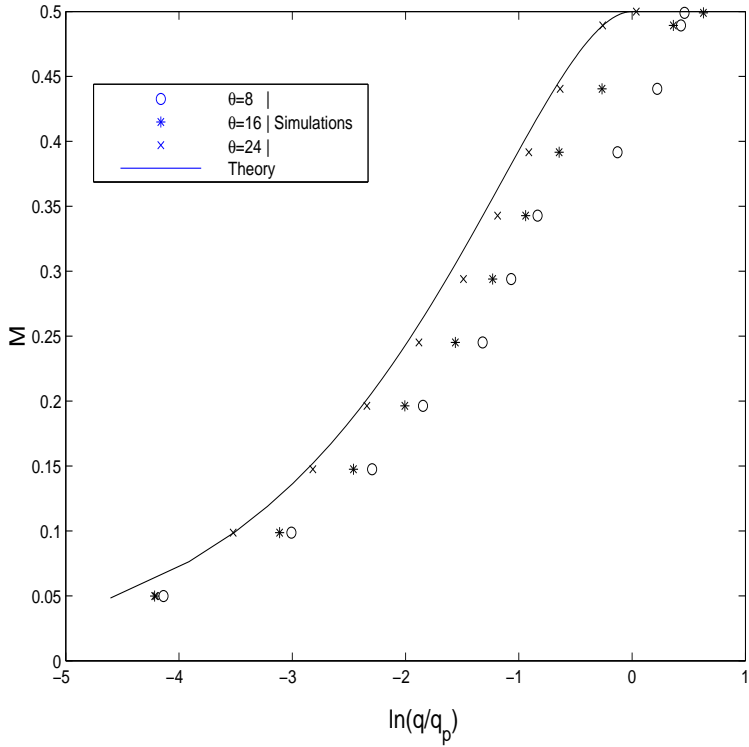,height=10truecm,width=10truecm}
\caption{\label{fig3a} \em Mass function $M(q, \theta)$ in lin-log scale
vs. normalized coordinate $q/q_p$. The particles are initially 
distributed uniformly in space, 
velocity a step-function of position.
The thick solid line represents the theoretical prediction
(\ref{massfunct1})($\gamma_{max}=1/2$). The circles, stars and crosses refer to the
simulation at different times.
}
\end{center}
\end{figure}

To validate our theoretical analysis we now turn to numerical simulations and 
consider the late stage of evolution of
an initial perturbation in the form (\ref{step}). The particles are
initially uniformly distributed in
space. In Figure \ref{fig3} the normalized density profile is 
plotted at late stage of evolution 
(thin solid line) and superposed to
the theoretical prediction (\ref{densstep}) (thick solid line).
The same curves are represented in scale log-log
in the left inset: here the circles refer to the simulation.
No parameter need to be adjusted by numerical fitting and the agreement
has to be considered satisfying.
The corresponding phase space portrait is reported 
in the right inset of Figure \ref{fig3}, where the
characteristic spiral behavior is clearly displayed. In 
figure \ref{fig3a} the cumulative mass function is plotted as a function
of the variable $\xi=q/q_p$.

\begin{figure}[ht!]
\begin{center}
\psfig{figure=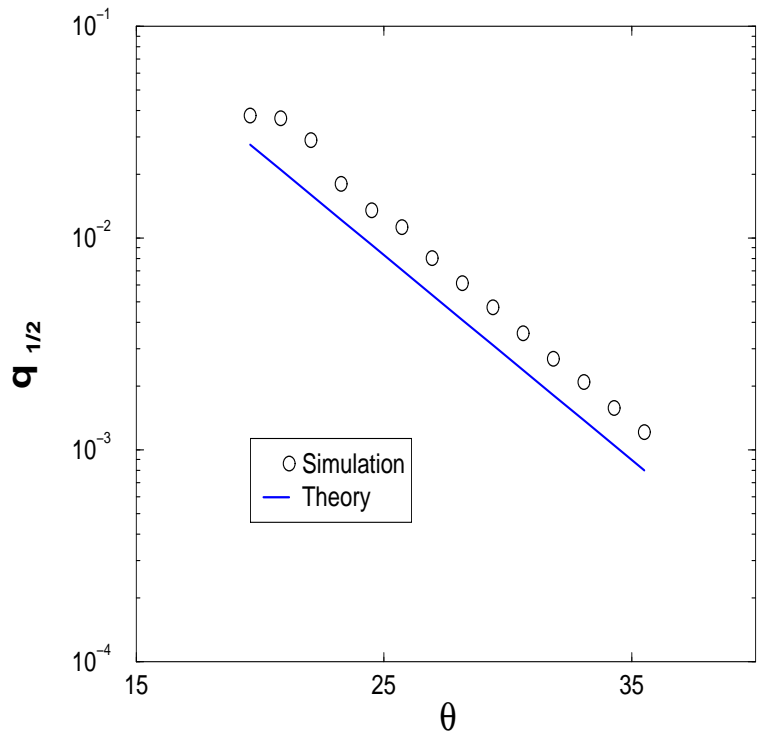,height=10truecm,width=10truecm}
\caption{\label{fig5} \em $q_{1/2}$ is plotted vs. $\theta$. The circles represent the results of
the numerical experiment. The solid line refers to the theoretical prediction
(equation (\ref{result})).}
\end{center}
\end{figure}

In a previous work of two of the authors, \cite{fanelli-aurell},
we studied the late time evolution of an
isolated perturbation within the framework of the Q dynamics.
In particular we measured the progressive
contraction of the inner region of the agglomeration, when compared
to the overall Universe expansion.
This effect was computed  by  the width of region, $\Delta q_{\mu}$, that
contains a fraction $\mu$ of the whole mass of the system, centered around the position
of maximum density. It turned out that the interval $\Delta q_{\mu}$ shrinks in time
according to a power-law
of cosmological time $t$, with exponent $-2/9$. An heuristic interpretation 
was also provided in \cite{fanelli-aurell}.
To make a bridge between our present investigations and these earlier findings 
we define $q_{\mu}$ (\ref{massfuncself})
such that $2 M(q_{\mu}, \theta)=\rho_0 \mu=const$ . 
From equation (\ref{mass-late}) it follows that  $\Delta q_{\mu}$ scales 
proportionally to $q_p$; thus we are led to assume  $q_{\mu}=x_{\mu}q_p$, 
where $x_{\mu}$ is a positive coefficient.
Equation(\ref{widthstep}) implies:
\begin{equation}
\label{result}
q_\mu= \frac{ \chi x_{\mu} } {2^{\frac{4}{3}} \beta_0^{\frac{1}{3}}}  \exp \left( - \frac{2}{9} \theta \right)
\end{equation}
in qualitative agreement with the results in \cite{fanelli-aurell}. As an example, 
let us focus on the case $\mu=1/2$. In this case,
$x_{\mu}=0.14$. We performed numerical simulations starting with the same class of initial
condition as in Figure \ref{fig3}. 
In Figure \ref{fig5}, $q_{1/2}$ is plotted vs. $\theta$:
circles refer to the numerical simulation and the solid line represents equation (\ref{result}). 
The slopes of the two curves in figure \ref{fig5} agree, 
but the amplitudes do not. We can make
the difference smaller by adjusting some of the 
hypotheses made in the derivation above. In particular, we considered
the particles belonging to the cluster to be uniformly distributed in the finite 
interval $\Delta q_{max} $ given by (\ref{deltaq_max}).
In reality, the particles spend more time on the border. This effect is 
explicitly modeled by assuming:

\begin{equation}
q_{p,eff}=c q_p
\end{equation}
where $c$ is an ``ad hoc'' numerical factor, larger than one.

To complete this section, let us compare our theoretical expression (\ref{densstep})
with the density profile derived in the framework of
the adhesion model. These results, which we recall in the following,
are detailed in \cite{gurbatov-density}. Let as consider the stationary  solution of 
Burgers equation :
\begin{equation}
\label{vadhstat}
v(x,b)= v_{st}(x)= -U \tanh \left( \frac{x}{\delta} \right),
\end{equation}
where ${\delta}=U/2 \nu$ is the width of the shock. This solution is also the asymptotic
solution of Burgers equation for the initial the step-function
profile previously considered.  
The trajectories of individual particles $x(b)$
satisfy the following equation
\begin{equation}
\label{part}
{\frac{dx(b)}{db}}=v(x,b),
\end{equation}
where the field $v(x,b)$ is determined by the solution of the
Burgers equation.  Assuming that $v(x,b)$ is a stationary solution 
of Burgers equation  from (\ref{vadhstat}),(\ref{part}) we have the
following expression for the coordinates of the particles
\begin{equation}
\label{bur4}
x(b,y)=\delta \hbox{Arc}\sinh \left[ \sinh (y/\delta )\exp (-U b/\delta)\right]
\end{equation}
where $y$ stands for the Lagrangian coordinate. From the conservation of mass, 
the following expression for the density is derived:
\begin{equation}
\label{rooldpan}
\rho(x,b) = \frac {\rho_0 \cosh \left( \frac{x}{\delta}\right) } { \sqrt{ \sinh^2 \left( \frac{x}{\delta}\right) + \exp \left( \frac{-2Ub}{\delta}\right) } },
\end{equation}
Solutions (\ref{bur4}) do not apply to the intermediate
stage of evolution. Thus we will focus on  
the asymptotic behavior, for time $b \gg \delta/U$
and consider the regions $|x|>\delta_p$, where $\delta_p =  \delta  \exp (-Ub/ \delta)$.
The particles in the agglomeration behave according to
\begin{equation}
\label{buras}
x(b,y)=y\exp (-U b/\delta).
\end{equation}
Thus in the adhesion model a faster collapse is displayed, when compared to the 
exact Q dynamics: in the latter a power-law decay in ``Burgers time'' $b$, is in 
fact produced, namely $q_p(b,y)=y (b/b_0)^{-1/3}$.
As concern the density profile, for $x<\delta_p$, one gets:
\begin{equation}
\label{rooldpan_as}
\rho(x,b) = \frac {\rho_0 \exp \left( \frac{U b}{\delta}\right) } { \sqrt{ 1 + \left(  \frac{x^2}{\delta^2} \right) \exp \left( \frac{2Ub}{\delta}\right) } },
\end{equation}
hence maximum value of the density increases exponentially
in ``Burgers time'',
$\rho_{max}(0,b) =  \rho_0 \exp ( Ub/\delta )$.
In the interval  $\delta_p \ll x \ll \delta$  the density transforms into a
time-independent power-law distribution
\begin{equation}
\label{rooldpow}
\rho(x,b) = \rho_0 \frac{\delta}{x}=\rho_0 \frac{2 \nu}{ U x}.
\end{equation}
Thus in the adhesion model we have that the asymptotic density 
distribution for the initial step profile is localized in the 
regions $|x|<\delta =const$, has non-integrable time independent 
power law tails (\ref{rooldpow}) with cutoff scale 
$\delta_p =  \delta  \exp (-Ub/ \delta)$.
In the Q model we have instead an integrable power-law distribution, see 
(\ref{densstep}). 

When considering a perturbation in a finite box, in co-moving coordinates, 
both the Quintic and the adhesion models show the formation of a dense structure.
However, in the adhesion model, the width $q_p$ of the pancake
decreases faster ( $g_p \sim \exp {-b}$) then in Q-model ( $g_p \sim b^{-1/3}$ ).
\\\\
These discrepancies suggest that, at least in one dimension, the adhesion 
approach is valid only as an approximate model of structure formation. This 
observation agrees with the conclusion in \cite{fanelli-aurell} where a new 
transport equation was proposed.


\section{Interaction of two massive clusters}
\label{s:2clust}

In this section we consider the dynamics of two interacting clusters.
As a first approximation, one can neglect the inner discrete structure
of a cluster, by introducing a single {\it macro-particle} to mimic its
dynamics. Thus, the problem is first reduced to the study of the
evolution of a two particles system, investigated in Section~\ref{s:2party}.
According to this picture, the agglomerations move back and forth
displaying damped oscillation. This observation agrees with
previous numerical investigations reported in \cite{fanelli}.
Following the discussion of the previous section,
each massive agglomeration experiences a contraction of the inner core:
the competition of these two effects determines the 
late-time asymptotics of the system.

\begin{figure}[ht!]
\begin{center}
\psfig{figure=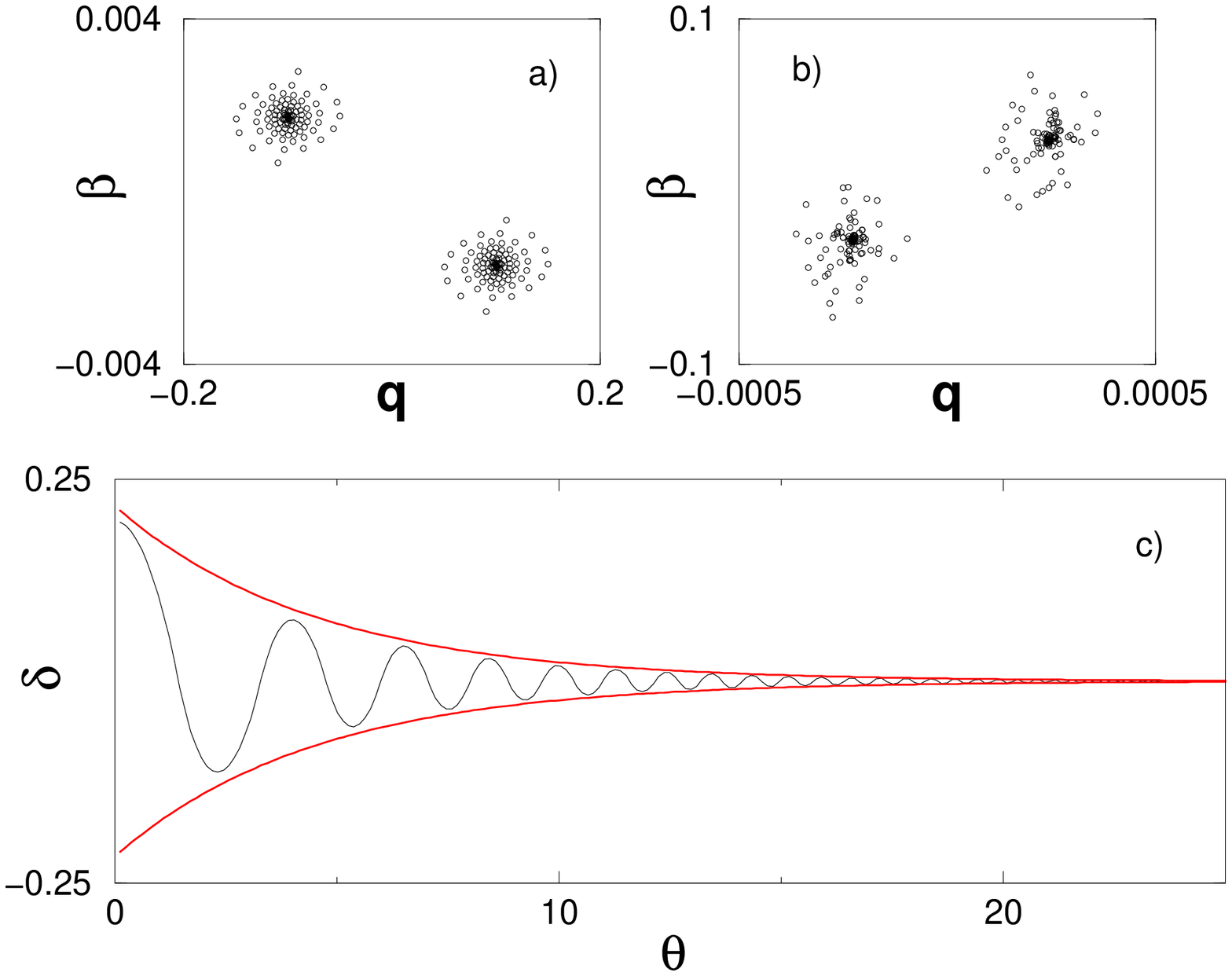,height=10truecm,width=10truecm}
\caption{\label{fig6} \em Figure a): Initial phase space portraits. Here
$N=200$. Figure b): Phase space portrait at time $\theta=25.5$.
Figure c):  $\delta$ vs. rescaled time $\theta$. The thin solid line
refers to the simulation. The thick curves represent eq. (\ref{deltaq_osc}).
Here $\theta_{cross}=1.38$.}
\end{center}
\end{figure}

Numerical simulation are performed starting with the initial condition
displayed in Figure \ref{fig6} a). The distance between the two centers of mass, $\delta$,
is plotted vs. time $\theta$, in Figure \ref{fig6} c). The damping tendency is evident
(thin solid line) and the theoretical prediction (\ref{deltaq_osc})
(thick solid lines) agrees with the numerical findings. 
At late time the two clusters still present a finite
separation, see \ref{fig6} b). This is in a good agreement with
the theoretical predictions of  Sections \ref{s:2party} and  \ref{sb:stepprofile},
both the widths of the agglomerations  and amplitude of their mutual 
oscillations decay as the same exponential function of time, hence their  
ratio is a constant number. Therefore, the system composed by two  
interacting clusters, with finite inner structure, evolves 
approximately in a self-similar way.

\section{Concluding remarks}
\label{s:con}
In this paper  we considered the problem of structure formation in a
Einstein-de Sitter Universe, focusing on planar perturbations.
This was done by using the Q model, or Quintic model, a Lagrangian
representation previously
derived in \cite{Qmodel}, in which one of the steps is to solve a large
number of quintic equations.
We underline however that the Q model is nothing but the motion
of a system of gravitationally interacting particles, on the background
of an expanding Universe, with convenient choices of time and space
coordinates.
We showed that the Q model can be approximated by an 
even simpler iterative
map. This observation permitted to establish a connection with
\cite{GuMaiTaMo} where the  motion of turbulent flows with strong friction was
analyzed, and to derive theoretical predictions for both the density profile
and the mass function 
in a collapsing cluster.
These predictions were cross-validated with direct numerical simulation.
\\\\
The main conclusion of this work is that the inner structure of a Zeldovich
pancake is different from the adhesion model, with finite viscosity in
the Burgers equation. That there is a discrepancy is what one would expect,
but this had
nevertheless to our knowledge never been shown before. 
As a mathematical problem, a qualitative difference is that 
in co-moving coordinates,
the size of collapsing cluster goes down as power-law with cosmological
time in the self-gravitating system, but as a stretched exponential
in the adhesion model
~\footnote{In ``Burgers time'' the collapse is
simply exponential, see subsection~\protect\ref{sb:stepprofile}.}.
As a physical problem, the difference is even more dramatic:
in the self-gravitating system
the characteristic scale of the density distribution is given
in co-moving coordinates by equation~(\ref{widthstep}), which
decreases, with cosmological time $t$, as $\left(t/t_0\right)^{-\frac{2}{9}}$.
The co-moving system of coordinates is related to an inertial
system by equation~(\ref{transf}),
where the proportionality is $\left(t/t_0\right)^{\frac{2}{3}}$.
Combining the two, the characteristic scale of a collapsing cluster
hence {\it increases} in time, as $d(t)\sim\left(t/t_0\right)^{\frac{4}{9}}$.
Since the density distribution of equation (\ref{densstep}) is
self-similar, the maximum density attained
depends on the graininess of
the initial conditions. 
We have not investigated the law of the increase of the maximum
density, but it is at least possible that with 
continuous initial density distributions arbitrarily
high densities can be produced, also if measured in an inertial
system of reference.
A more interesting characteristics
is however the fraction of total mass 
$M$ where (in an inertial system)
the density is higher than some cut-off $\rho_{cr}$.
That quantity scales as $\left(\frac{M}{\rho_{cr} d(t)}\right)^3$.
Since $d(t)$ is an increasing
function of time in the inertial system,
that fraction decreases as 
$\left(t/t_0\right)^{-\frac{4}{3}}$. 
In this sense one-dimensional self-gravitating motion,
on the background of an expanding Universe, is not effective
in bringing about mass collapse.
\\\\
In the adhesion model the  characteristic scale decreases, both
in a co-moving and an inertial system. Here both the 
maximum real-space density and the fraction of mass at a density
larger than some cut-off \textit{increases}. The adhesion
model is therefore, in contrast, effective in bringing about mass
collapse in one dimension.
In other words, the adhesion model
over-predicts the clustering tendency, and overestimates the peak density.
A Universe in which most mass lies in regions where
density decreases can never
give rise to galaxies, galaxy clusters and other 
significant objects. The analysis given here therefore
suggests, quantitatively, why a one-dimensional description
of mass agglomeration is insufficient.
\\\\
We have also investigated the mutual motion of two pancakes, and shown
them to be closely similar to the decaying oscillations of two
mutually attracting ``heavy particles''.


\section{Acknowledgements}
This work was mainly supported by a grant from the Swedish Royal Academy of
Sciences (KVA), further support from
the Swedish Research Council through grants
NFR F~650-19981250 (D.F) and NFR I~510-930 (E.A.), and
RFBR (02-02-17374, 00-15-96619) and "Universities of Russia" (S.N.G. and A.Yu.M) is also
gratefully acknowledged.
E.A. thanks the Edwin Schr\"odinger Institute (Vienna) for hospitality,
and for opportunities to discuss and write up this work. D.F. thanks Ph. Choquard
for enlighting discussions. We thank  A.~Noullez for fruitful discussions.




\newpage

\begin{thebibliography}{99}


\bibitem{Qmodel} Aurell~E. \& Fanelli~D., cond-mat/0106444 (2001)

\bibitem{BOOMERANG} \textit{BOOMERANG Home Page},
http:$//$www.physics.ucsb.edu/$\sim$boomerang/~.

\bibitem{Buchert1} Buchert~T. \& Dom\'inguez~A., {\it A\&A} ,  {\bf 335}, 395 (1999).

\bibitem{Buchert2} Buchert~T., Dom\'inguez~A. \& Perez-M\'ercader,
  {\it A\&A} , {\bf 349}, 343 (1999).

\bibitem{cal} Caldwell~R.R., Dave~R. \& Steinhardt~P.J.,
\textit{Phys. Rev. Lett.} \textbf{80} 1582 (2001).

\bibitem{COBE-web-site} \textit{ COBE Home Page},
        http:$//$www.gsfc.nasa.gov/astro/cobe/ 
/cobe$\_$home.html.


\bibitem{DMR-web-site} \textit{ DMR images}
                This web site includes
                qualitative color-coded pictures of the anisotropies
                in the Cosmic Background Radiation field as observed by the
COBE satellite, as well as down-loadable data sets,
                http:$//$www.gsfc.nasa.gov/astro/cobe/dmr$\_$image.html.


\bibitem{DOROSHKEVICH}
Doroshkevich~A.G., Kotok~E.V., Novikov~I.D., Poludov~A.N., Shandarin~S.F. \&  Si
gov~Yu.S. at al. 1980, {\it MNRAS},{\bf 192}, 321


\bibitem{fanelli}  Fanelli~D., TRITA-NA-0209, NADA, KTH, Stockholm, Sweden

\bibitem{fanelli-aurell} Fanelli~D., Aurell~E.,   {\it Astronomy \& Astrophysics} 
(2002) [in press].

\bibitem{japan} Fanelli~D., Aurell~E. \& Noullez~A.,
                Proceeding of IAU Symposium 208 (2001).

\bibitem{gurbatov-density} Gurbatov S.N., Proceedings of the International school of physics
``E. Fermi'', Course CXXXII, Dark Matter in the Universe, Societ\'a Italiana di Fisica,645-660,1996.

\bibitem{libro}
 Gurbatov S.N.,   Malakhov A.N.,  Saichev A.I. {\it Nonlinear Random
Waves and Turbulence in Nondispersive Media: Waves, Rays, Particles}
(Manchester University Press, 1991).

\bibitem{GuMaiTaMo}
Gurbatov S.N., Mainardi F., Moshkov A.Yu., Tampieri F.
Izvestia VUZov:PND, Saratov, 4-5 (2002)


\bibitem{Gurbatov} Gurbatov~S.N., Saichev~A.I. \&
                Shandarin~S.F. 1989, {\it MNRAS}, {\bf 236}, 385.

\bibitem{Harrison1970} Harrison~E.R., \textit{ Phys. Rev. D} \textbf{ 1} (1970) 2726-2730.


\bibitem{KamionkowskiKosowsky} Kamionkowsky~M. \& Kosowsky~A.,
                 \textit{Ann. Rev. Nucl. Part. Sci.} \textbf{49} (1999) 77-123.


\bibitem{Lifshitz} Lifshitz~E.,  J. Phys. USSR, {\bf 10}, 116 (1947).

\bibitem{MAP-web-site} \textit{MAP Home Page},
http:$//$map.gsfc.nasa.gov/~.


\bibitem{al}    Noullez~A., Fanelli~D. \& Aurell~E.,
                submitted to Journ. Comp. Phys.,  cond-mat/0101336 (2001).


\bibitem{Peebles1980}  Peebles~P.J. , The Large-scale
                Structure of the Universe,
                (Princeton University Press, Princeton, NJ) (1980).

\bibitem{num}   Press,~H.W., Numerical Recipes in Fortran,
                (Cambridge University Press, Cambridge) (1992).


\bibitem{Rouet1990} Rouet~J.L., Feix~M.R. \& Navet M.,
                Vistas in Astronomy, {\bf 33}, 357 (1990).


\bibitem{Rouet1991} Rouet~J.L. et al.,
                in Lecture Notes in Physics:
                Applying Fractals in Astronomy, 161 (1991).


\bibitem{Zeldovich} Shandarin~S.F. \& Zeldovich~Ya.B.,
                Rev. Mod. Phys., {\bf 61}, 185 (1989).


\bibitem{Tegmark-science} Tegmark~M., 
{\it Science} {\bf 296} 1427-1433 (2002).

\bibitem{YANO-GOUDA}
Yano~T.\& N.Gouda~N., {\it The astrophysical Journal. Suppliment Series}
{\bf 118} 267 (1998).




\bibitem{Vergassola} Vergassola~M., Dubrulle~B.,
        Frisch~U. \& Noullez~A., {\it A\&A}, {\bf 289}, 325 (1993).

\bibitem{Weinberg1972} Weinberg~S., Gravitation and
        Cosmology (Wiley) (1972).

\bibitem{Wein1} Weinberg~S.,
\textit{Phys. Rev. D} \textbf{64} (2001) 123511.

\bibitem{Wein2} Weinberg~S.,
\textit{Phys. Rev. D} \textbf{64} (2001) 123512.

\end{thebibliography}
\end{document}